\documentclass[11pt]{article}
\usepackage{amsfonts}
\newcommand{\eq}{\begin{equation}}
\newcommand{\en}{\end{equation}}
\newcommand{\eqn}{\begin{eqnarray}}
\newcommand{\enn}{\end{eqnarray}}

\newcommand{\beq}{\begin{equation}}
\newcommand{\eeq}{\end{equation}}
\newcommand{\tn}{\ensuremath{\tilde{n}}}
\newcommand{\ta}{\ensuremath{\tilde{a}}}
\newcommand{\tb}{\ensuremath{\tilde{b}}}
\newcommand{\tc}{\ensuremath{\tilde{c}}}
\newcommand{\tx}{\ensuremath{\tilde{x}}}
\newcommand{\ty}{\ensuremath{\tilde{y}}}
\newcommand{\ti}{\ensuremath{\tilde{I}}}
\newcommand{\tj}{\ensuremath{\tilde{J}}}
\newcommand{\tk}{\ensuremath{\tilde{K}}}

\newcommand{\bi}{\ensuremath{\bar{i}}}
\newcommand{\bj}{\ensuremath{\bar{j}}}
\newcommand{\bk}{\ensuremath{\bar{k}}}
\newcommand  {\Rbar} {{\mbox{\rm$\mbox{I}\!\mbox{R}$}}}
\newcommand  {\Hbar} {{\mbox{\rm$\mbox{I}\!\mbox{H}$}}}
\newcommand {\Cbar}{\mathord{\setlength{\unitlength}{1em}
     \begin{picture}(0.6,0.7)(-0.1,0) \put(-0.1,0){\rm C}
        \thicklines \put(0.2,0.05){\line(0,1){0.55}}\end {picture}}}
\begin{document}
\begin{titlepage}
\begin{flushright}
  PSU-TH-222\\
%CERN-TH/99-???\\
\end{flushright}
\vspace{0.2cm}
\begin{center}
\begin{LARGE}
\textbf{The Gauging of Five-dimensional, 
$\mathcal{N}=2$ Maxwell-Einstein Supergravity Theories
Coupled to Tensor Multiplets}\footnote{ Work supported in part by the National
Science Foundation under Grant Number PHY-9802510.}
\end{LARGE}\\
\vspace{1.0cm}
\begin{large}
M. G\"{u}naydin$^{\dagger\ddagger}$ \footnote{murat@phys.psu.edu} and
M. Zagermann$^{\ddagger}$ \end{large}\footnote{zagerman@phys.psu.edu}  \\
\vspace{.35cm}
$^{\dagger}$ CERN, Theory Division \\
1211 Geneva 23, Switzerland \\
and \\
$^{\ddagger}$ Physics Department \\
Penn State University\\
University Park, PA 16802, USA \\
\vspace{0.5cm}
{\bf Abstract} 
\end{center}
\begin{small}
We study the general gaugings of $\mathcal{N}=2$ Maxwell-Einstein
supergravity theories (MESGT) in five dimensions, extending and generalizing
previous work. The global symmetries 
of these theories are of the form $SU(2)_R \times G$,
 where $SU(2)_R$ is the 
$R$-symmetry group of the $\mathcal{N}=2$ Poincar\'{e}
 superalgebra and $G$ is the 
group of isometries of the scalar manifold that extend to symmetries of 
the full
action. We first gauge  a subgroup $K$ of $G$ by turning some of the
vector fields into gauge fields of $K$ while dualizing the 
remaining vector fields
into tensor fields transforming in a non-trivial representation of $K$. 
 Surprisingly, we find that
the presence of tensor fields transforming non-trivially under the Yang-Mills gauge
group leads to the introduction of a potential which does not admit an AdS ground state.
Next we give the simultaneous gauging of the $U(1)_R$ subgroup of $SU(2)_R$ and a subgroup
$K$ of $G$ in the presence of $K$-charged tensor multiplets. 
The potential introduced by the simultaneous
 gauging is the sum of the potentials introduced by gauging $K$ and $U(1)_R$ separately.
We present a list of possible gauge groups $K$ and the corresponding representations
of tensor fields. For the exceptional supergravity we find that one can gauge the
 $SO^*(6)$ subgroup of the isometry group $E_{6(-26)}$ of the scalar manifold 
if one dualizes 12 of the vector fields to tensor fields just as in the gauged 
 $\mathcal{N}=8$
supergravity.

\end{small}

\end{titlepage}

\renewcommand{\theequation}{\arabic{section}.\arabic{equation}}
\section{Introduction}
\setcounter{equation}{0}

Gauged supergravity\footnote{
The term ``gauged supergravity'' commonly refers to (usually  
$\mathcal{N}$-extended) supergravity theories in which a subgroup of the  
automorphism group (alias ``R-symmetry group'') of the underlying 
supersymmetry algebra is realized as a local (Yang-Mills-type) gauge 
symmetry. Sometimes, this term is also used for gaugings of other 
global
symmetry groups that are not subgroups of the 
R-symmetry group. In this paper, we will refer to the latter 
type of theories as ``Yang-Mills/Einstein supergravity theories''. 
In contrast, ``ungauged'' supergravity theories are those 
for which the
R-symmetry group is just a global symmetry group of the Lagrangian.}
theories in various dimensions  have been studied
 extensively in the early and mid-eighties (see e.g. 
\cite{dWN,PPvN7,GRW0,GRW1,PPvN,SS}).

In the last few years there has been a renewed intense interest 
in  gauged
supergravity theories. This interest is driven mainly by the work
on AdS/CFT (anti-de Sitter/conformal field theory)  dualities \cite{jm,
GKP,EW98,earlier,agmoo}.   For example, the
  IIB superstring theory on the background manifold
$AdS_{5}\times S^{5}$ with $N$ units of five-form flux through the five-sphere,
is conjectured to be equivalent (at least in a certain limit) 
to $4d$ $\mathcal{N}=4$ super
 Yang Mills theory with gauge
group $SU(N)$, which is a conformally invariant quantum field theory.
In the limit of small string coupling and large $N$, the classical 
(ie. tree level) IIB supergravity approximation becomes valid and can be 
used to discuss the large $N$ limit of the corresponding dual Yang Mills 
theory. The importance of \emph{gauged} supergravity lies 
in the fact that $5d$ 
gauged $\mathcal{N}=8$  supergravity \cite{GRW0,GRW1,PPvN}
 is believed to be a consistent nonlinear
truncation of the lowest lying Kaluza Klein modes of IIB supergravity on
$AdS_{5}\times S^{5}$ \cite{GM,krv}.\footnote{ The consistency of the nonlinear truncation of the 
$S^7$ and $S^4$  
compactifications of 11-dimensional supergravity was shown in \cite{dWN2} and \cite{NVvN},
respectively.} 
 Many aspects of the AdS/CFT correspondence, such as the
 renormalization group flows of 
certain non-conformal deformations of the Yang Mills theory with a
smaller
number of supersymmetries, can therefore
 be studied entirely within the framework of $5d$ gauged supergravity 
due to the lack of interference with the higher Kaluza-Klein modes 
\cite{FGPW,npw}.
 Thus, gauged supergravity theories 
lie at the core of AdS/CFT dualities.
 
On the other hand,  five-dimensional $\mathcal{N}=2$ gauged
 supergravity is the natural 
framework for so-called brane world scenarios in which our $4d$ world
is realized as a domain wall in an  effectively five-dimensional   
theory \cite{RS1,RS2,BC}. In fact,
certain M-theory compactifications \cite{HW1,EW2,LOSW1,ELPP} 
seem to suggest
theories which appear five-dimensional at a certain intermediate
length scale, at which the effective field theory is given by a
certain $5d$  $\mathcal{N}=2$ gauged supergravity plus $4d$ Standard 
Model-type matter fields on the $4d$ boundaries of this $5d$ spacetime.

Motivated by the above-mentioned applications, as well as others, we 
study the most general gaugings of $5d$,  $\mathcal{N}=2$ supergravity
theories coupled to vector as well as {\it tensor} multiplets. The work
presented here represents a generalization and an extension of earlier
work on the gaugings of $\mathcal{N}=2$ supergravity coupled to 
vector multiplets
\cite{GST1,GST2,GST3,GST4,GST5}. 

The organization of the paper is as 
follows. For the convenience of the reader, section 2 briefly summarizes
the basic features of \emph{ungauged} Maxwell-Einstein supergravity
theories. Focussing on the global symmetries of these ungauged theories,
we list the possible types of their gaugings. The subsequent four sections
describe each of these gauge types in detail: Section 3 summarizes the
gauging of a $U(1)_{R}$ subgroup of the $\mathcal{N}=2$ R-symmetry group
$SU(2)_R$. Sections 4 and 5 are devoted to the gauging of a
subgroup $K$ of the isometry group $G$ of the scalar manifold: Section 4
summarizes the well-known case without tensor fields, whereas Section 5
covers the case when tensor fields have to be introduced. The
simultaneous gauging of $U(1)_{R}$ and $K$ is treated in Section 6. We conclude
with a  classification of possible gauge groups and the corresponding 
representations of tensor fields in Section 7 and a short
discussion of our results in Section 8.

\section{Ungauged $\mathcal{N}=2$ Maxwell-Einstein supergravity theories 
 and their global symmetries}
\setcounter{equation}{0}
In this section, we briefly recall the most  relevant features
of the  (ungauged) $\mathcal{N}=2$ 
Maxwell-Einstein supergravity theories (MESGT) constructed in \cite{GST1}.
Unless otherwise stated, our conventions will coincide
with those of ref. 
\cite{GST1}, where further details can be found. 
In particular, we will use the  metric signature 
$(-++++)$ and impose the `symplectic' Majorana condition on 
all fermionic quantities.

The fields of the $\mathcal{N} =2$ supergravity multiplet are the
f\"{u}nfbein $e_{\mu}^{m}$, two gravitini $\Psi_{\mu}^{i}$ ($i=1,2$)
and a vector field $A_{\mu}$. An $\mathcal{N} =2$ vector multiplet contains
a vector field $A_{\mu}$, two spin-$1/2$ fermions $\lambda^{i}$ and one real 
scalar field $\varphi$. The fermions of each of these multiplets transform as 
doublets under the $USp(2)_{R}\cong SU(2)_{R}$ R-symmetry 
group of the 
$\mathcal{N} =2$ Poincar\'{e} superalgebra; all other fields are 
$SU(2)_{R}$-inert.

The  $\mathcal{N}=2$ MESGT's constructed in \cite{GST1}
describe the coupling of $\tn$ vector multiplets to supergravity.
Hence, the total field content is
\begin{equation}
\{ e_{\mu}^{m}, \Psi_{\mu}^{i}, A_{\mu}^{\ti}, \lambda^{i\ta}, \varphi^{\tx}\}
\end{equation}
with
\begin{eqnarray*}
\ti&=& 0,1,\ldots, \tn\\
\ta&=& 1,\ldots, \tn\\
\tx&=& 1,\ldots, \tn,
\end{eqnarray*}
where we have combined the `graviphoton' with the $\tn$ vector fields 
of the $\tn$ vector multiplets into a single $(\tn+1)$-plet of vector fields 
$A_{\mu}^{\ti}$ labelled by the index $\ti$.   
The indices $\ta, \tb, \ldots$ and $\tx, \ty, \ldots$ should be interpreted as
flat and curved indices, respectively, of
the
 $\tn$-dimensional target space manifold $\mathcal{M}$ of the scalar fields.
(Our indices $(\ti,\ta,\tx)$ correspond to the indices $(I,a,x)$ in refs. 
\cite{GST1,GST2,GST3}.)

The generic Maxwell-Einstein supergravity Lagrangian was found to be
(up to 4-fermion terms) \cite{GST1}:
\begin{eqnarray}\label{Lagrange}
e^{-1}\mathcal{L}&=& -\frac{1}{2}R(\omega)-\frac{1}{2}
{\bar{\Psi}}_{\mu}^{i}\Gamma^{\mu\nu\rho}\nabla_{\nu}\Psi_{\rho i}-
\frac{1}{4}{\stackrel{\circ}{a}}_{\ti\tj}F_{\mu\nu}^{\ti}
F^{\tj\mu\nu}\nonumber\\
&& -\frac{1}{2}{\bar{\lambda}}^{i\ta}\left(\Gamma^{\mu}\nabla_{\mu}
\delta^{\ta\tb}+
\Omega_{\tx}^{\ta\tb}\Gamma^{\mu}\partial_{\mu}\varphi^{\tx}\right)
\lambda_{i}^{\tb}-\frac{1}{2}g_{\tx\ty}(\partial_{\mu}\varphi^{\tx})
(\partial^{\mu}
\varphi^{\ty})\nonumber\\
&&-\frac{i}{2}{\bar{\lambda}}^{i\ta}\Gamma^{\mu}\Gamma^{\nu}\Psi_{\mu i}
f_{\tx}^{\ta}\partial_{\nu}\varphi^{\tx}+ \frac{1}{4}h_{\ti}^{\ta}
{\bar{\lambda}}^{i\ta}\Gamma^{\mu}\Gamma^{\lambda\rho}\Psi_{\mu i}
F_{\lambda\rho}^{\ti}
\nonumber\cr
&&+\frac{i}{2\sqrt{6}}\left(\frac{1}{4}\delta_{\ta\tb}h_{\ti}+T_{\ta\tb\tc}
h_{\ti}^{\tc}\right)
{\bar{\lambda}}^{i\ta}\Gamma^{\mu\nu}\lambda_{i}^{\tb}
F_{\mu\nu}^{\ti}\nonumber\\
&&-\frac{3i}{8\sqrt{6}}h_{\ti}\left[{\bar{\Psi}}_{\mu}^{i}
\Gamma^{\mu\nu\rho\sigma}
\Psi_{\nu i}F_{\rho\sigma}^{\ti}+2{\bar{\Psi}}^{\mu i}
\Psi_{i}^{\nu}F_{\mu\nu}^{\ti}\right]%\nonumber
\cr
&& + \frac{e^{-1}}{6\sqrt{6}}C_{\ti\tj\tk}\varepsilon^{\mu\nu\rho\sigma\lambda}
 F_{\mu\nu}^{\ti}F_{\rho\sigma}^{\tj}A_{\lambda}^{\tk}
\end{eqnarray}
with the supersymmetry transformation laws (to leading order in fermion fields)
\begin{eqnarray}
\delta e_{\mu}^{m}&=& \frac{1}{2}{\bar{\varepsilon}}^{i}
\Gamma^{m}\Psi_{\mu i}\nonumber\cr
\delta \Psi_{\mu i} &=&\nabla_{\mu} (\omega)\varepsilon_{i}+\frac{i
}
{4\sqrt{6}}h_{\ti}(\Gamma_{\mu}^{\:\:\:\nu\rho}-4\delta_{\mu}^{\nu}
\Gamma^{\rho})F_{\nu\rho}^{\ti}\varepsilon_{i}\nonumber\cr
\delta A_{\mu}^{\ti}&=& \vartheta_{\mu}^{\ti}\nonumber\cr
\delta \lambda_{i}^{\ta}  &=& -\frac{i}{2}f_{\tx}^{\ta}
\Gamma^{\mu}(\partial_{\mu}\varphi^{\tx})\varepsilon_{i} + 
\frac{1}{4}h_{\ti}^{\ta}
\Gamma^{\mu\nu}
\varepsilon_{i}F_{\mu\nu}^{\ti}\\
\delta \varphi^{\tx}&=&\frac{i}{2}f^{\tx}_{\ta}{\bar{\varepsilon}}^{i}
\lambda_{i}^{\ta},\label{trafo}
\end{eqnarray}
where
\begin{equation}
\vartheta_{\mu}^{\ti}\equiv -\frac{1}{2}h_{\ta}^{\ti}{\bar{\varepsilon}}^{i}
\Gamma_{\mu}\lambda_{i}^{\ta}+\frac{i\sqrt{6}}{4}h^{\ti}
{\bar{\Psi}}_{\mu}^{i}\varepsilon_{i}.
\end{equation}
Here, $e$ denotes the f\"{u}nfbein determinant, whereas $R(\omega)$
and $\nabla_{\mu}=\nabla_{\mu}(\omega)$ are the scalar
curvature and the spacetime covariant derivative with respect 
to the ordinary spin connection $\omega_{\mu}^{mn}(e)$. 
$F_{\mu\nu}^{\ti}$ are the field strengths of the Abelian vector
fields $A_{\mu}^{\ti}$. 
The various scalar field dependent quantities that contract the 
different types of indices are as follows: $f_{\tx}^{\ta}$, $g_{\tx\ty}$
and $\Omega_{\tx}^{\ta\tb}$ denote the $\tn$-bein, the metric  and the 
spin connection, 
respectively, of the target manifold $\mathcal{M}$. The quantities 
$h_{\ti}$, $h^{\ti}$, $h_{\ti}^{\ta}$,
$h^{\ti}_{\ta}$, $T_{\ta\tb\tc}$ and ${\stackrel{\circ}{a}}_{\ti\tj}$
are $\varphi^{\tx}$-dependent functions that are subject to various 
algebraic and differential constraints  (see \cite{GST1} for details)   
as required by supersymmetry.
These constraints also involve $f_{\tx}^{\ta}$, $g_{\tx\ty}$,
$\Omega_{\tx}^{\ta\tb}$ and imply that \emph{all} scalar field
dependent quantities are completely determined by the \emph{constant}
symmetric tensor $C_{\ti\tj\tk}$ that appears in the $F\wedge F\wedge A$-
term in (\ref{Lagrange}). The $C_{\ti\tj\tk}$ thus uniquely
determine the whole theory.
In particular, the scalar field target manifold $\mathcal{M}$ can 
be viewed as an $\tn$-dimensional hypersurface 
\begin{equation}\label{hyper}
C_{\ti\tj\tk}h^{\ti}h^{\tj}h^{\tk}=1.
\end{equation}
of an (\tn+1)-dimensional ambient space parametrized by  (\tn+1)
coordinates $h^{\ti}$. The resulting geometry of these theories  was later 
referred to as  
``very special geometry''. In \cite{GST3} it was suggested that the 
compactification of 11-dimensional supergravity over a Calabi-Yau threefold
would lead to $d=5$ , $\mathcal{N}=2$ MESGT's coupled to hypermultiplets.
 The Calabi-Yau 
compactifications of $11d$ supergravity  were later studied in \cite{ferrara1}
where it was explicitly shown that they lead to $\mathcal{N}=2$ MESGT's with
$(h_{(1,1)}-1)$ vector multiplets coupled to $(h_{(2,1)}+1)$ hypermultiplets.
($h_{(1,1)}$ and $h_{(2,1)}$ are the Hodge numbers of the corresponding
Calabi-Yau manifold.)  

The $C_{\ti\tj\tk}$ themselves are not completely arbitrary.
Going     to a particular  basis \cite{GST1}, they can be brought 
to the following form
\begin{equation}\label{canbasis}
C_{000}=1,\quad C_{0ij}=-\frac{1}{2}\delta_{ij},\quad  C_{00i}=0
\end{equation}
and  the remaining coefficients $C_{ijk}$
 ($i,j,k=1,2,\ldots , \tn$) may be chosen 
at will. We shall refer to this basis as the canonical basis. 

The arbitrariness of the  $C_{ijk}$ shows that, even for a fixed
number $\tn$ of vector multiplets, various target manifolds
$\mathcal{M}$  are possible. A classification of these ``very special real''
manifolds has been given in \cite{dWvP1} for the case that 
 $\mathcal{M}$ is a homogeneous space. This class contains the subclass
of symmetric spaces, which were  classified already long time ago
\cite{GST1,GST3}. Although our further discussion is not at all restricted to
symmetric (or even homogeneous) $\mathcal{M}$, we will look at the 
symmetric spaces in a somewhat greater detail in section 7. Let us therefore 
list the possible symmetric spaces for later reference.
The symmetric spaces $\mathcal{M}$ fall into two
different categories, depending on whether  they are associated with 
 Jordan algebras or not:\\
(i) $\mathcal{M}= \frac{\textrm{Str}_{0}(\textrm{J})}{\textrm{Aut}
(\textrm{J})}$, where 
$\textrm{Str}_{0}$(J) 
and 
Aut(J) are the reduced structure group and the automorphism group,
respectively, of a formally real, unital, Jordan algebra, J, of degree
three \cite{GST1, KMC}. This ``Jordan class'' can be further divided into two 
subclasses:

\begin{itemize}
\item ``Generic'' or ``reducible'' Jordan class:\\
\eq 
\textrm{J}=\mathbf{R}\oplus \Sigma_{\tn}: \qquad \mathcal{M} = \
\frac{\textrm{SO}(\tn-1,1)\times \textrm{SO}(1,1)}{\textrm{SO}(\tn-1)}, 
\qquad \tn\geq 1.\nonumber
\en
Here, $\Sigma_{\tn}$ is a Jordan algebra of degree two, which can be identified
 as the algebra of Dirac gamma matrices in an $(\tn-1)$-dimensional
(internal) `Minkowski' space with the product 
being one half the anticommutator. 
\item ``Irreducible'' or ``magical'' Jordan class. The corresponding 
Jordan algebras are simple and are isomorphic to the Hermitian 
$(3\times 3)$-matrices 
over the four division algebras $\mathbf{R}, \mathbf{C}, \mathbf{H}, 
\mathbf{O}$ with the product being the anticommutator. They  lead to the 
following target spaces:
\begin{eqnarray}
\textrm{J}_{3}^{\mathbf{R}}:\quad \mathcal{M}&=& \textrm{SL}(3,\mathbf{R})/
\textrm{SO}(3),\qquad
(\tn=5)\nonumber\cr
\textrm{J}_{3}^{\mathbf{C}}:\quad \mathcal{M}&=& \textrm{SL}(3,\mathbf{C})/
\textrm{SU}(3),\qquad
(\tn=8)\nonumber\cr
\textrm{J}_{3}^{\mathbf{H}}:\quad \mathcal{M}&=& \textrm{SU}^{*}(6)/
\textrm{Usp}(6),\qquad
(\tn=14)\nonumber\cr
\textrm{J}_{3}^{\mathbf{O}}:\quad \mathcal{M}&=& \textrm{E}_{6(-26)}/
\textrm{F}_{4},\qquad \qquad
(\tn=26)\nonumber
\end{eqnarray}

\end{itemize}
(ii)  $\mathcal{M}=\frac{\textrm{SO}(1,\tn)}{\textrm{SO}(\tn)},\quad \tn>1$. 
This class is not 
associated with  Jordan algebras and will therefore be referred to as 
the ``symmetric non-Jordan-family'' \cite{GST3}.

We will now turn to the global symmetries of a generic  MESGT (with 
possibly non-symmetric or non-homogeneous $\mathcal{M}$) 
described by (\ref{Lagrange}). Two different global symmetries have to be 
distinguished:
\begin{itemize}
\item Any $\mathcal{N}=2$ MESGT is globally invariant under the R-symmetry 
group $SU(2)_{R}$. This symmetry is inherited from the underlying 
supersymmetry algebra and acts exclusively on the fermions $\Psi_{\mu}^{i}$
and $\lambda^{i\ta}$ (ie. on their index $i$).
\item Any group G of linear transformations
\begin{displaymath}
h^{\ti}\rightarrow {B^{\ti}}_{\tj}h^{\tj}, \quad A_{\mu}^{\ti}\rightarrow 
{B^{\ti}}_{\tj}A_{\mu}^{\tj}
\end{displaymath} that leaves the 
tensor $C_{\ti\tj\tk}$ invariant
\begin{displaymath}
{B^{\ti'}}_{\ti}{B^{\tj'}}_{\tj}{B^{\tk'}}_{\tk}C_{\ti'\tj'\tk'}=C_{\ti\tj\tk}
\end{displaymath}
is automatically a symmetry of the whole Lagrangian (\ref{Lagrange}),
since the latter is uniquely  determined by 
the $C_{\ti\tj\tk}$.
In particular, these symmetries give rise to isometries of the scalar
manifolds $\mathcal{M}$, which becomes manifest if one rewrites the kinetic
energy term for the scalar fields as \cite{GST1,dWvP1}
\begin{displaymath}
-\frac{1}{2}g_{\tx\ty}(\partial_{\mu}\varphi^{\tx})(\partial^{\mu}
\varphi^{\ty})= \frac{3}{2}C_{\ti\tj\tk}h^{\ti}\partial_{\mu}h^{\tj}
\partial^{\mu}h^{\tk}
\end{displaymath}
with the $h^{\ti}$ constrained according to (\ref{hyper}).
\end{itemize}

Important (but not the only) examples with such a  non-trivial symmetry group 
$G$ are given by the aforementioned symmetric space cases. In the Jordan
class, $G$ coincides with the full isometry group of $\mathcal{M}$
(ie. with the full ``numerator group'' $\textrm{Str}_{0}(\textrm{J})$). 
For the symmetric non-Jordan 
family, $G=[SO(\tn-1)\times SO(1,1)]\odot T_{(\tn-1)}$
where $\odot$ denotes the semi-direct product and $T_{(\tn -1)}$ is the
group of translations in an 
$(\tn -1)$ dimensional Euclidean space. Note that 
for this family $G$ is  only a subgroup of the target 
space isometry group $\textrm{SO}(1,\tn)$ \cite{dWvP2}.

The fact that the total global symmetry group of (\ref{Lagrange}) factorizes
into $SU(2)_{R}\times G$ is a consequence of the  $SU(2)_{R}$-invariance
of the scalar fields belonging to the vector multiplets and
 allows to study the  gaugings
of the two factors separately. In general 
matter coupled extended supergravity theories the R-symmetry group is 
nontrivially embedded into some larger global symmetry group if the scalar
 fields are not singlets under it.

Let us now turn to the possible gaugings of subgroups of $SU(2)_{R}\times G$.
Since the vector fields are all $SU(2)_{R}$-inert, they cannot  
 serve
as non-Abelian gauge fields for the full $SU(2)_{R}$\footnote{One 
could, however, try to identify $SU(2)_{R}$ with an $SU(2)$-subgroup of $G$ 
and then 
gauge this diagonal subgroup, yet  we have not considered such a possibility
in the present paper.}.  We will therefore only consider gaugings of subgroups
of $U(1)_{R}\times G$, where $U(1)_{R}$ denotes the $U(1)$ subgroup of 
$SU(2)_{R}$. This obviously leaves the following possibilities:
\begin{enumerate}
\item
 One can simply gauge the $U(1)_{R}$ subgroup of $SU(2)_R$
 by coupling a linear combination 
of the 
vector fields to the fermions \cite{GST2}, which are the only fields that 
transform nontrivially under $SU(2)_{R}$. In general, this kind of gauging 
(which we will refer to as ``gauged MESGT'') introduces a scalar potential
(see Section 3). 
\item Another possibility is to gauge a subgroup $K$ of $G$. In this 
case, 
which 
we will refer to as ``Yang-Mills/Einstein supergravity'', at least a 
subset of the 
vector fields has to transform in the adjoint representation of $K$ so that
 these vector fields
can serve as the corresponding Yang-Mills gauge fields. If there are additional vector 
fields (`spectator vector fields') beyond these gauge fields, there are two 
possibilities. They are either
$K$-singlets  or some of them transform non-trivially under $K$. In the former
case, there are no technical difficulties and the gauging can be performed as 
described in \cite{GST2} and leads to a theory \emph{without} scalar 
potential (see Section 4)    
\item If there \emph{are}  vector fields that are charged under $K$, one 
faces the same 
problem that was first encountered in the context of maximally extended 
gauged 
supergravity in  seven \cite{PPvN7} and subsequently  in five dimensions
\cite{GRW0,GRW1,PPvN}.  
The problem is  that a naive 
gauging of $K$ would introduce masses for these vector fields, 
thereby leading to a mismatch between bosonic and fermionic
degrees of freedom. The only known solution to this problem is to convert
the charged vector fields into two-form fields with ``self-dual'' field 
equations \cite{PTvN}. In the maximally extended theories, this idea is also 
supported by the analysis of the spectra of the underlying Kaluza-Klein 
compactifications \cite{GM,krv}. 
\item Finally, one can combine (i) and (ii), or alternatively (i) and (iii), 
and 
simultaneously gauge both $U(1)_{R}$ and $K\subset G$. 
We will 
refer to this 
type of gauging as ``gauged Yang-Mills/Einstein supergravity''. 
\end{enumerate}

The first two possibilities were studied in \cite{GST2,GST4,GST5} with special 
emphasis on the cases were $\mathcal{M}$ is a symmetric space of the Jordan 
family.
It is the purpose of this paper to extend some of the aspects that were 
discussed in \cite{GST2,GST4,GST5} for the gaugings of type (i) and (ii)
 to more general 
$\mathcal{M}$ and, moreover, 
study the so far uncovered gaugings of type (iii) and (iv),
thereby closing a gap in the existing literature.

\section{Gauged Maxwell/Einstein supergravity}
\setcounter{equation}{0}
In order to gauge the $U(1)_{R}$-subgroup of the $SU(2)_{R}$ R-symmetry group,
one promotes a linear combination of the $(\tn+1)$ vector fields 
$A_{\mu}^{\ti}$
to the $U(1)_{R}$-gauge field
\begin{equation}\label{AV}
A_{\mu}[U(1)_{R}]=V_{\ti}A_{\mu}^{\ti},
\end{equation}
where $V_{\ti}$ are $(\tn+1)$ constants, and replaces the derivatives of 
the fermionic fields by $U(1)_{R}$-covariant derivatives
\begin{eqnarray}
\nabla_{\mu}\lambda^{\ta i} &\longrightarrow& (D_{\mu}\lambda^{\ta })^{i}
\equiv  \nabla_{\mu}\lambda^{\ta i}+g_{R}V_{\ti}A_{\mu}^{\ti}\delta^{ij}
\lambda_{j}^{\ta}\nonumber\\
\nabla_{\mu}\Psi_{\nu}^{i}&\longrightarrow& (D_{\mu}\Psi_{\nu})^{i}
\equiv \nabla_{\mu}\Psi_{\nu}^{i}+g_{R}V_{\ti}A_{\mu}^{\ti}\delta^{ij}\Psi_{
\nu j},
\end{eqnarray}
where $g_{R}$ denotes the $U(1)_{R}$-coupling constant. The appearance
of the $\delta^{ij}$ is due to the convention that the 
$SU(2)_{R}$-indices $i,j,\ldots$ are raised and lowered with the 
antisymmetric
metric $\varepsilon_{ij}=-\varepsilon_{ji}$, 
$\varepsilon_{12}=\varepsilon^{12}=1$ \cite{GST1,  GST2}.
This $U(1)_{R}$-covariantization in the Lagrangian (\ref{Lagrange}) and 
the transformation laws (\ref{trafo}) breaks the original supersymmetry.
In order to restore it, some  $g_{R}$-dependent gauge invariant terms 
have to be added. The additional terms in the Lagrangian are \cite{GST2,GST5}
(the numerical factors are chosen for convenience)
\begin{eqnarray}
e^{-1}\mathcal{L}' &=&-\frac{i\sqrt{6}}{8} g_{R} {\bar{\Psi}}_{\mu}^{i}
\Gamma^{\mu\nu}
\Psi_{\nu}^{j}\delta_{ij} P_{0}(\varphi)-\frac{1}{\sqrt{2}}
g_{R}{\bar{\lambda}}^{i\ta}
\Gamma^{\mu}
\Psi_{\mu}^{j}\delta_{ij} P_{\ta}(\varphi)\nonumber\\
&&+\frac{i}{2\sqrt{6}}g_{R}{\bar{\lambda}}^{i\ta}
\lambda^{j\tb}\delta_{ij} P_{\ta\tb}(\varphi)-g_{R}^{2}P^{(R)}
(\varphi),
\end{eqnarray}
whereas the transformation laws have to be modified by
\begin{eqnarray}
\delta' \Psi_{\mu}^{i}&=&\frac{i}{2\sqrt{6}}g_{R}P_{0}(\varphi)\Gamma_{\mu}
\delta^{ij}\varepsilon_{j}\nonumber\\
\delta' \lambda^{i\ta}&=& \frac{1}{\sqrt{2}}g_{R}P^{\ta}
(\varphi)\delta^{ij}\varepsilon_{j}.
\end{eqnarray}
The new scalar field dependent quantities $P_{0}$, $P^{\ta}$, $P_{\ta\tb}$,
and the scalar potential  $P^{(R)}$ are fixed by supersymmetry
\begin{eqnarray}\label{Pcons} 
P^{\ta}&=&\sqrt{2}h^{\ta \ti}V_{\ti}\label{Pcons1}\\
P_{0}&=&2 h^{\ti}V_{\ti}\label{Pcons2}\\
P_{\ta\tb}&=&\frac{1}{2}\delta_{\ta\tb}P_{0}+2\sqrt{2}
T_{\ta\tb\tc}P^{\tc}\label{Pcons3}\\
P^{(R)}&=&-(P_{0})^{2}+P^{\ta}P^{\ta}\label{Pcons4}.
\end{eqnarray}
The scalar potential $P^{(R)}$ can be written in the form \cite{GST2,GST5}
\begin{equation}
P^{(R)}= -4 C^{\ti \tj \tk} V_{\ti}V_{\tj}h_{\tk},
\end{equation}
where the $\ti, \tj, \tk$ are raised with the inverse of 
${\stackrel{\circ}{a}}_{\ti\tj}$.
In our metric signature, a critical point $\varphi_{c}$ 
of the scalar potential with $P^{(R)}(\varphi_{c})<0$ 
corresponds to an anti-de Sitter  ground state. 
The critical points of the potential (\ref{Pcons4})
have been analyzed in \cite{GST2,GST5} for 
the Jordan cases. If a critical point exists, it was found that, 
depending on the linear combination (\ref{AV}) of 
the vector fields, one either gets  an $\mathcal{N}=2$ supersymmetric 
anti-de Sitter ground state, or the scalar potential vanishes identically, 
and thus admits a Minkowski vacuum with spontaneously broken 
supersymmetry.

\section{Yang-Mills/Einstein supergravity without tensor fields}
\setcounter{equation}{0}
We now consider the gauging of a subgroup $K$ of $G$. As mentioned earlier,
this type requires that a subset $\{A_{\mu}^{I};\quad  I,J,\ldots =1,\ldots 
\textrm{dim}K\}$
of  the vector fields transforms in the adjoint representation of $K$.
In this section, we assume that if there are additional 
spectator vector fields
$\{A_{\mu}^{M};\quad M,N,P=1,\ldots,(\tn+1)-\textrm{dim}{K}\}$, they are all 
$K$-singlets (i.e., we are dealing with the gauging of type (ii)).
 
The only fields that transform under $K$ are the scalar fields 
$\varphi^{\tx}$,
the spinor fields $\lambda^{i\ta}$ and the vector fiels $A_{\mu}^{I}$, 
($I=1,\ldots \textrm{dim} K$). 
The $K$-covariantization is thus achieved by replacing the corresponding
derivatives/field strenghts by their $K$-gauge covariant counterparts:
\begin{eqnarray}
\partial_{\mu}\varphi^{\tx}&\longrightarrow & \mathcal{D}_{\mu}
\varphi^{\tx}\equiv
\partial_{\mu}\varphi^{\tx}+gA_{\mu}^{I}K_{I}^{\tx}\nonumber\\
\nabla_{\mu}\lambda^{i\ta}&\longrightarrow & \mathcal{D}_{\mu}
\lambda^{i\ta}\equiv
\nabla_{\mu}\lambda^{i\ta} +gA_{\mu}^{I} L_{I}^{\ta\tb}\lambda^{i\tb}
\nonumber\\
F_{\mu\nu}^{I}&\longrightarrow &\mathcal{F}_{\mu\nu}^{I}\equiv 
F_{\mu\nu}^{I}+gf_{JK}^{I}A_{\mu}^{J}A_{\nu}^{K}.
\end{eqnarray}
Here, $g$ denotes the $K$-coupling constant, 
$K_{I}^{\tx}$ are the Killing vectors of 
$\mathcal{M}$ that correspond to the subgroup K of its isometry group G
(cf. \cite{GST2}), $L_{I}^{\ta\tb}$ are the (scalar field dependent) 
K-transformation 
matrices
of the fermions $\lambda^{i\ta}$ (cf. \cite{GST2,GST4}) and $ f_{JK}^{I}$ 
are the structure constants of K.
These replacements in the Lagrangian (\ref{Lagrange}) and the 
transformation laws  (\ref{trafo}) are subject to one exception:
The proper gauge-covariantization of the $F\wedge F\wedge A$-term in 
(\ref{Lagrange})
leads to a Chern Simons term, i.e., $\frac{e^{-1}}{6\sqrt{6}}
C_{\ti\tj\tk}\varepsilon^{\mu\nu\rho\sigma\lambda}
 F_{\mu\nu}^{\ti}F_{\rho\sigma}^{\tj}A_{\lambda}^{\tk}$ gets replaced by
\begin{eqnarray}
\frac{e^{-1}}{6\sqrt{6}}C_{\ti\tj\tk}\varepsilon^{\mu\nu\rho\sigma\lambda}
\left\{ F_{\mu\nu}^{\ti}F_{\rho\sigma}^{\tj}A_{\lambda}^{\tk} + \frac{3}{2}g
F_{\mu\nu}^{\ti}A_{\rho}^{\tj}(f_{\tilde{L}\tilde{M}}^{\tk}
A_{\sigma}^{\tilde{L}}A_{\lambda}^{\tilde{M}}) \right.
+\left. \frac{3}{5}g^{2}(f_{\tilde{N}\tilde{P}}^{
\tj}
A_{\nu}^{\tilde{N}}A_{\rho}^{\tilde{P}})
(f_{\tilde{L}\tilde{M}}^{\tk}A_{\sigma}^{\tilde{L}}A_{\lambda}^{\tilde{M}})
A_{\mu}^{\ti}\right\},
\end{eqnarray}
where it is understood that $f_{\ti\tj}^{\tk}$ is zero whenever one of the
indices $\ti$, $\tj$, $\tk$ corresponds to one of the spectator vector
fields $A_{\mu}^{M}$. 

Again, supersymmetry is broken by these replacements. This time, however, 
its restauration requires  little modification; the 
(covariantized) transformation
laws remain unchanged, and only a Yukawa-like term has to be 
added to the (covariantized) Lagrangian \cite{GST2,GST4}
\begin{equation}
\mathcal{L}'= -\frac{i}{2}g{\bar{\lambda}}^{i\ta}\lambda_{i}^{\tb}K_{I[\ta}
h^{I}_{\tb]}.
\end{equation}
In particular, no scalar potential is introduced so that only Minkowski ground
states
are possible.

\section{Yang-Mills/Einstein supergravity with tensor fields}
\setcounter{equation}{0}
We now turn to case (iii) of our gauge type classification and consider
the gauging of $K\subset G$, when \emph{not} all the spectator vector 
fields are $K$-singlets. 
As mentioned earlier,  consistency 
with supersymmetry requires that
these $K$-charged spectator vector fields have to be dualized to 
 ``self-dual'' two-form fields \cite{PTvN,GRW0,GRW1,PPvN,GM,krv}.
 We will therefore split
the vector fields $A_{\mu}^{\ti}$ of the \emph{ungauged} theory 
(\ref{Lagrange})-(\ref{trafo}) of Section 2    into 
two sets. The first set contains the vector fields in the adjoint 
representation  of the 
gauge group $K$ plus possible $K$-singlets. The second set contains
the remaining \emph{$K$-charged} vector fields. 
We will use indices $I, J, K,\ldots
=1,\ldots, n$ for the first and $M,N,P,\ldots=1,\ldots 2m$ for the second set,
 where   $n+2m=\tn+1$. The reason for the even number $2m$ is  
that the ``self-duality''-condition of \cite{PTvN} requires 
\emph{complex} 
tensor fields for $d=5$, which we will always  consider 
 as  being decomposed  
into their 
real and 
imaginary parts.
The gauging now proceeds as follows.
First, one   has to replace all Abelian field strenghts $F_{\mu\nu}^{I}$ by
the corresponding non-Abelian generalizations 
$\mathcal{F}_{\mu\nu}^{I}\equiv 
F_{\mu\nu}^{I}+gf_{JK}^{I}A_{\mu}^{J}A_{\nu}^{K}$, with $ f_{JK}^{I}$ 
being the structure constants of $K$\footnote{In the presence of K-singlets, 
the corresponding $f_{JK}^{I}$ are again assumed to be zero (cf. Section 4)}, 
and the  $F_{\mu\nu}^{M}$
by the above-mentioned ``self-dual'' 
two-form fields $B_{\mu\nu}^{M}$:

\eq
F_{\mu\nu}^{\ti}\longrightarrow \mathcal{H}_{\mu\nu}^{\ti}
:=(\mathcal{F}_{\mu\nu}^{I},B_{\mu\nu}^{M}).
\en
Again, the only exception to this replacement is the $F\wedge F\wedge A$-term
of the ungauged theory. Since no `naked' $A_{\mu}^{M}$ can appear anymore,
  we first require
\begin{equation}
C_{MNP}=0
\end{equation}
and since terms of the form $B^{M}\wedge F^{I}\wedge A^{J}$ 
appear  to be impossible to supersymmetrize in a gauge invariant way
(except possibly in  very special cases) we shall also assume that
\begin{equation}
C_{MIJ}=0
\end{equation}
Hence, the only non-vanishing $C_{\ti\tj\tk}$ have the index structure
$C_{IJK}$ and $C_{IMN}$. The covariantization of the $C_{IJK}F^{I}\wedge F^{J}
\wedge A^{K}$-term again leads to a Chern-Simons term (see below).
The term of the form $C_{IMN}A^{I}\wedge B^{M}\wedge B^{N}$ has its natural 
place in the gauge-invariant kinetic energy term for the tensor fields 
$B_{\mu\nu}^{M}$  (cf. eqs.(\ref{DB}) and (\ref{Lagrange2})). 
The  gauge covariant derivative of these
tensor fields reads
\begin{equation}\label{DB}
\mathcal{D}_{\mu}B_{\nu\rho}^{M}\equiv\nabla_{\mu}B_{\nu\rho}^{M}+
gA_{\mu}^{I}\Lambda_{IN}^{M}
B_{\nu\rho}^{N},
\end{equation}
where the \emph{constant} matrices $\Lambda_{IN}^{M}$ are the corresponding
representation matrices of K.

The remaining gauge covariantizations involve the scalar and spinor fields,
for which we again make the replacements
\begin{eqnarray}
\partial_{\mu}\varphi^{\tx}&\longrightarrow & \mathcal{D}_{\mu}
\varphi^{\tx}\equiv
\partial_{\mu}\varphi^{\tx}+gA_{\mu}^{I}K_{I}^{\tx}\nonumber\\
\nabla_{\mu}\lambda^{i\ta}&\longrightarrow & \mathcal{D}_{\mu}\lambda^{i
\ta}\equiv
\nabla_{\mu}\lambda^{i\ta} +gA_{\mu}^{I} L_{I}^{\ta\tb}\lambda^{i\tb}.
\label{Dlambda}
\end{eqnarray}

After all these modifications, the original supersymmetry of
the ungauged theory (\ref{Lagrange})-(\ref{trafo}) is again
badly broken. This time, however, the supersymmetry breaking is not only 
due to the gauge covariantization alone. An additional source for
the breakdown of  supersymmetry  is provided by the loss of the 
Bianchi identity
for the tensor fields $B_{\mu\nu}^{M}$ (ie. $dB^{M}\neq0$ in general). 
The corresponding Bianchi identity $dF^{\ti}=0$
for the $F_{\mu\nu}^{\ti}$ in the ungauged theory is needed at several 
places to cancel certain supersymmetry variations in (\ref{Lagrange}).

Remarkably enough, supersymmetry can again be restored  by adding further 
$g$-dependent gauge invariant 
terms to the 
Lagrangian and the transformation laws. This procedure is very
similar to  what had to be  done in the 
$\mathcal{N}=8$ theory \cite{GRW0,GRW1,PPvN}. We omit the details here  
and quote the final result.

The Lagrangian is given by (up to 4-fermion terms)

\begin{eqnarray}\label{Lagrange2}
e^{-1}\mathcal{L}&=& -\frac{1}{2}R(\omega)-\frac{1}{2}
{\bar{\Psi}}_{\mu}^{i}\Gamma^{\mu\nu\rho}\nabla_{\nu}\Psi_{\rho i}-
\frac{1}{4}{\stackrel{\circ}{a}}_{\ti\tj}\mathcal{H}_{\mu\nu}^{\ti}
\mathcal{H}^{\tj\mu\nu}
\nonumber\cr
& & -\frac{1}{2}{\bar{\lambda}}^{i\ta}\left(\Gamma^{\mu}\mathcal{D}_{\mu}
\delta^{\ta\tb}+
\Omega_{\tx}^{\ta\tb}\Gamma^{\mu}\mathcal{D}_{\mu}\varphi^{\tx}\right)
\lambda_{i}^{\tb}-
\frac{1}{2}g_{\tx\ty}(\mathcal{D}_{\mu}\varphi^{\tx})(\mathcal{D}^{\mu}
\varphi^{\ty})\nonumber\cr
&& -\frac{i}{2}{\bar{\lambda}}^{i\ta}\Gamma^{\mu}\Gamma^{\nu}\Psi_{\mu i}
f_{\tx}^{\ta}\mathcal{D}_{\nu}\varphi^{\tx}+ \frac{1}{4}h_{\ti}^{\ta}
{\bar{\lambda}}^{i\ta}\Gamma^{\mu}\Gamma^{\lambda\rho}\Psi_{\mu i}
\mathcal{H}_{\lambda\rho}^{\ti}
\nonumber\cr
&&+\frac{i}{2\sqrt{6}}\left(\frac{1}{4}\delta_{\ta\tb}h_{\ti}+T_{\ta\tb\tc}
h_{\ti}^{\tc}\right)
{\bar{\lambda}}^{i\ta}\Gamma^{\mu\nu}\lambda_{i}^{\tb}
\mathcal{H}_{\mu\nu}^{\ti}\nonumber\cr
&& -\frac{3i}{8\sqrt{6}}h_{\ti}\left[{\bar{\Psi}}_{\mu}^{i}
\Gamma^{\mu\nu\rho\sigma}
\Psi_{\nu i}\mathcal{H}_{\rho\sigma}^{\ti}+2{\bar{\Psi}}^{\mu i}
\Psi_{i}^{\nu}\mathcal{H}_{\mu\nu}^{\ti}\right]\nonumber\cr
&& + \frac{e^{-1}}{6\sqrt{6}}C_{IJK}\varepsilon^{\mu\nu\rho\sigma\lambda}
\left\{ F_{\mu\nu}^{I}F_{\rho\sigma}^{J}A_{\lambda}^{K} + \frac{3}{2}g
F_{\mu\nu}^{I}A_{\rho}^{J}(f_{LF}^{K}A_{\sigma}^{L}A_{\lambda}^{F})\right. 
\nonumber\cr
&& \qquad\qquad\qquad\qquad+\left. 
\frac{3}{5}g^{2}(f_{GH}^{J}A_{\nu}^{G}A_{\rho}^{H})
(f_{LF}^{K}A_{\sigma}^{L}A_{\lambda}^{F})A_{\mu}^{I}\right\}\nonumber\cr
&&+\frac{e^{-1}}{4g}\varepsilon^{\mu\nu\rho\sigma\lambda}\Omega_{MN}
B_{\mu\nu}^{M}\mathcal{D}_{\rho}B_{\sigma\lambda}^{N}\nonumber\\
&&+g{\bar{\lambda}}^{i\ta}\Gamma^{\mu}\Psi_{\mu i}W^{\ta}+
g{\bar{\lambda}}^{i\ta}\lambda_{i}^{\tb}W^{\ta\tb}-g^{2}P.%\label{Lagrange}
\end{eqnarray}
The transformation laws are (to leading order in fermion fields)
\begin{eqnarray}\label{trafo2}
\delta e_{\mu}^{m}&=& \frac{1}{2}{\bar{\varepsilon}}^{i}
\Gamma^{m}\Psi_{\mu i}\nonumber\cr
\delta \Psi_{\mu i} &=&\nabla_{\mu} (\omega)\varepsilon_{i}+\frac{i
}
{4\sqrt{6}}h_{\ti}(\Gamma_{\mu}^{\:\:\:\nu\rho}-4\delta_{\mu}^{\nu}
\Gamma^{\rho})\mathcal{H}_{\nu\rho}^{\ti}\varepsilon_{i}\nonumber\cr
\delta A_{\mu}^{I}&=& \vartheta_{\mu}^{I}\nonumber\cr
\delta B_{\mu\nu}^{M} &=& 2\mathcal{D}_{[\mu}\vartheta_{\nu]}^{M} +
\frac{\sqrt{6}g}{4}
\Omega^{MN}h_{N}{\bar{\Psi}}^{i}_{[\mu}\Gamma_{\nu]}\varepsilon_{i}
+\frac{ig}{4}\Omega^{MN}h_{N}^{\ta}{\bar{\lambda}}^{i\ta}\Gamma_{\mu\nu}
\varepsilon_{i}\nonumber\\
\delta \lambda_{i}^{\ta}  &=& -\frac{i}{2}f_{\tx}^{\ta}
\Gamma^{\mu}(\mathcal{D}_{\mu}
\varphi^{\tx})\varepsilon_{i} + \frac{1}{4}h_{\ti}^{\ta}
\Gamma^{\mu\nu}
\varepsilon_{i}\mathcal{H}_{\mu\nu}^{\ti}+gW^{\ta}\varepsilon_{i}\nonumber\\
\delta \varphi^{\tx}&=&\frac{i}{2}f^{\tx}_{\ta}{\bar{\varepsilon}}^{i}
\lambda_{i}^{\ta}
\end{eqnarray}
with
\begin{equation}
\vartheta_{\mu}^{\ti}\equiv -\frac{1}{2}h_{\ta}^{\ti}{\bar{\varepsilon}}^{i}
\Gamma_{\mu}\lambda_{i}^{\ta}+\frac{i\sqrt{6}}{4}h^{\ti}
{\bar{\Psi}}_{\mu}^{i}\varepsilon_{i}.
\end{equation}

The quantities which are not already present in the ungauged theory are
a (constant) real symplectic metric $\Omega_{MN}$
\begin{equation}
\Omega_{MN}=-\Omega_{NM}, \qquad \Omega_{MN}\Omega^{NP}=\delta_{M}^{P},
\end{equation}
two tensors $W^{\ta}(\varphi)$ and $W^{\ta\tb}(\varphi)$
\begin{eqnarray}
W^{\ta}&=&-\frac{\sqrt{6}}{8}h_{M}^{\ta}\Omega^{MN}h_{N}\nonumber\\
W^{\ta\tb}&=&-W^{\tb\ta}= ih_{\, }^{J[\ta}K_{J}^{\tb]}+\frac{i\sqrt{6}}{4}
h^{J}K_{J}^{\ta;\tb},
\end{eqnarray}
where the semicolon denotes covariant differentiation on the target space 
$\mathcal{M}$, and a scalar potential $P(\varphi)$
\begin{equation}\label{P}
P=2W^{\ta}W^{\ta}.
\end{equation}
Furthermore, one finds the relation 
\begin{equation}
\Lambda_{IM}^{N}=\frac{2}{\sqrt{6}}\Omega^{NP}C_{MPI} \Longleftrightarrow
\Omega_{NP}\Lambda_{IM}^{P}=\frac{2}{\sqrt{6}} C_{MNI},
\end{equation}
which, because of $C_{MNI}=C_{NMI}$, means that the 
$\Lambda_{IM}^{N}$ have to form a symplectic 
representation of the gauge group $K$.
Supersymmetry also requires the following two relations
\begin{eqnarray}
W^{\ta}&=&\frac{\sqrt{6}}{4}h^{J}K_{J}^{\ta}\\
P^{,\ta}&=& 4i W^{\ta\tb}W^{\tb}
\end{eqnarray}
where the comma denotes partial differentiation with respect to 
the scalar fields. These last two  conditions, however, 
can be shown to follow automatically from the various other constraints.

The above scalar potential 
$P(\varphi)$ deserves some comments.

First of all, it is a  bit surprising that there is a scalar potential 
at all, since no minimal couplings to the gravitini have been introduced at 
this point, 
and, as we have seen in Section 4,   the pure 
Yang-Mills/Einstein 
supergravity theories \emph{without} antisymmetric tensor fields do 
\emph{not} involve a scalar potential. In fact, the necessity for the scalar 
potential in the above Lagrangian can eventually be traced back to the
loss of the Bianchi identity for the $B_{\mu\nu}^{M}$, which are not present 
in the theories considered in Section 4. 

The second important point about the potential is its sign. 
As mentioned at the end of Section 3, in our  metric signature, 
a critical point with $P(\varphi_{c})<0$ 
would correspond to an Anti-de Sitter solution. The explicit form (\ref{P}) 
of our  potential, however, is manifestly non-negative. Therefore, 
the $\mathcal{N}=2$ Yang-Mills/Einstein supergravity theories with 
tensor multiplets do \emph{not} admit  an Anti-de Sitter solution. 
This might at first seem surprising, since it was, among other things, 
the representation theory of the $AdS_{5}$-superalgebra $SU(2,2|4)$ that
hinted towards the dualization of twelve  vector fields to antisymmetric 
tensor fields in the gauging of the $\mathcal{N}=8$ supergravity theory
 in $d=5$
\cite{GRW0,GRW1,PPvN}.
 For $\mathcal{N}=2$ however, this argument does not apply anymore, 
since the $\mathcal{N}=2$ Anti-de Sitter graviton supermultiplet also contains
only one vector field and no tensor fields, giving rise to the same field 
content as  its  $\mathcal{N}=2$ super Poincar\'{e} counterpart. Thus,
for $\mathcal{N}=2$, the antisymmetric tensor fields do not necessarily 
have to be associated with 
Anti-de Sitter spacetimes anymore.

\section{Gauged Yang-Mills/Einstein supergravity with tensor \\
 fields}
\setcounter{equation}{0}
We will now come to   case (iv) of our list of possible
gaugings and simultaneously gauge the $U(1)_R$ R-symmetry subgroup 
and 
a  subgroup $K$ of $G$. We will do this for the most general case with tensor
fields, since the case without tensor fields can easily be recovered as
 a special case.
Our starting point will be the Yang-Mills/Einstein supergravity with tensor 
fields presented in the previous section, i.e. eqs.  
(\ref{Lagrange2})-(\ref{trafo2}). 

As in Section 3, we will take  a linear combination of the vector fields 
$A_{\mu}^{I}$ as the $U(1)_R$-gauge field
\begin{equation}\label{AU1}
A_{\mu}{[}U(1)_R{]} = V_{I}A_{\mu}^{I}
\end{equation}
with some constants $V_{I}$, which at this point are completely arbitrary.
(Note, however, that we don't sum over $\ti$ like in Section 3, 
i.e., ``$V_{M}=0$''.) 
The gauging of  $U(1)_R$ then obviously requires the  
$U(1)_R$-covariantization of all fermionic derivatives:
\begin{eqnarray}
\mathcal{D}_{\mu}\lambda^{\ta i} &\longrightarrow& (\mathfrak{D}_{\mu}
\lambda^{\ta })^{i}
\equiv  \mathcal{D}_{\mu}\lambda^{\ta i}+g_{R}V_{I}A_{\mu}^{I}\delta^{ij}
\lambda_{j}^{\ta}\nonumber\\
\nabla_{\mu}\Psi_{\nu}^{i}&\longrightarrow& (\mathfrak{D}_{\mu}\Psi_{\nu})^{i}
\equiv \nabla_{\mu}\Psi_{\nu}^{i}+g_{R}V_{I}A_{\mu}^{I}\delta^{ij}\Psi_{\nu j},
\end{eqnarray}
where $g_{R}$ again denotes the $U(1)_R$-coupling constant and $\mathcal{D}_{\mu}$
is the $K$-covariant derivative introduced in (\ref{Dlambda}). 
Again, this gauge covariantization breaks supersymmetry, and to restore it, 
new
$g_{R}$-dependent terms have to be added to the Lagrangian and 
the transformation laws.  

The additional terms in the transformation laws are
\begin{eqnarray}
e^{-1}\mathcal{L}' &=&-\frac{i\sqrt{6}}{8} g_{R} {\bar{\Psi}}_{\mu}^{i}
\Gamma^{\mu\nu}
\Psi_{\nu}^{j}\delta_{ij} P_{0}(\varphi)-\frac{1}{\sqrt{2}}
g_{R}{\bar{\lambda}}^{i\ta}
\Gamma^{\mu}
\Psi_{\mu}^{j}\delta_{ij} P_{\ta}(\varphi)\nonumber\\
&&+\frac{i}{2\sqrt{6}}g_{R}{\bar{\lambda}}^{i\ta}
\lambda^{j\tb}\delta_{ij} P_{\ta\tb}(\varphi)-g_{R}^{2}P^{(R)}
(\varphi),
\end{eqnarray}
whereas the transformation laws have to be modified by
\begin{eqnarray}
\delta' \Psi_{\mu}^{i}&=&\frac{i}{2\sqrt{6}}g_{R}P_{0}(\varphi)\Gamma_{\mu}
\delta^{ij}\varepsilon_{j}\nonumber\\
\delta' \lambda^{i\ta}&=& \frac{1}{\sqrt{2}}g_{R}P^{\ta}
(\varphi)\delta^{ij}\varepsilon_{j}.
\end{eqnarray}
The new scalar field dependent quantities $P_{0}$, $P^{\ta}$, $P_{\ta\tb}$,
and the scalar potential  $P^{(R)}$ are fixed by supersymmetry
\begin{eqnarray}\label{U1cons} 
P^{\ta}&=&\sqrt{2}h^{\ta I}V_{I}\label{U1cons1}\\
P_{0}&=&2 h^{I}V_{I}\label{U1cons2}\\
P_{\ta\tb}&=&\frac{1}{2}\delta_{\ta\tb}P_{0}+2\sqrt{2}
T_{\ta\tb\tc}P^{\tc}\label{U1cons3}\\
P^{(R)}&=&-(P_{0})^{2}+P^{\ta}P^{\ta}\label{U1cons4}.
\end{eqnarray}
Furthermore, the $V_{I}$ are constrained by
\begin{equation}\label{Vf}
V_{I}f_{JK}^{I}=0.
\end{equation}
Supersymmetry also requires the relations
\begin{eqnarray}
P^{\ta}K^{\ta}_{I}&=&0\nonumber\\
P_{\ta\tb}W^{\tb}&=&-i 2\sqrt{3}W_{\ta\tb}
P^{\tb}+\frac{5}{2}W_{\ta}P_{0}\nonumber\\
P^{\ta}_{;\tx}&=&\frac{-\sqrt{3}}{4}P_{0}f_{\tx}^{\ta}-\frac{1}{2\sqrt{3}}
P_{\ta\tb}
f_{\tx}^{\tb}\nonumber\\
P_{0,\tx}&=&-\frac{2}{\sqrt{3}}P^{\ta}f_{\tx}^{\ta}\nonumber\\
P^{(R)}_{,\tx}&=&\frac{5}{2\sqrt{3}}P_{0}P^{\ta}
f_{\tx}^{\ta}-\frac{1}{\sqrt{3}}f_{\tx}^{\ta}
P_{\ta\tb}P^{\tb},
\end{eqnarray}
however, these can be shown to be consequences of the other constraints
and therefore do not give rise to additional restrictions.

It should be noted that the constraints (\ref{U1cons1})-(\ref{U1cons4}) 
are  almost the  same  as
in the case of the pure 
$U(1)_R$-gauging described in Section 3. 
Yet there are two important differences. The first  is 
that the (completely arbitrary) $V_{\ti}$ of Section 3 are now 
subject to two constraints, namely  eq. (\ref{Vf}) and ``$V_{M}=0$'', which 
is merely a trivial 
consequence of (\ref{AU1}).

The second difference is that (\ref{U1cons4}) is not the full scalar potential.
The latter is 
 now a sum of the 
$U(1)_R$-related potential $P^{(R)}$ and the potential 
$P$, which  was due to
 the 
introduction of the 2-form fields (cf. eqs. (\ref{Lagrange2}) and (\ref{P})):
\begin{equation}
e^{-1}\mathcal{L}_{pot}=-g^{2}P-g_{R}^{2}P^{(R)}
\end{equation}
These differences have some interesting implications:

Eq. (\ref{Vf}) gives a new constraint on the possible gauge groups $K$, 
since for it to be true, the $f_{JK}^{I}$ have to admit a nontrivial 
eigenvector $V_{I}$ with eigenvalue 0. This means that either there has to be 
at least one spectator vector field $A_{\mu}^{I}$ or       K has to have at 
least one Abelian factor (both of them together could   also be true).

As for the potential, one sees       that the $U(1)_R$-gauging 
introduces
a negative contribution to the total scalar potential 
so that Anti-de Sitter solutions might now be possible. In fact, the experience
with the gauged MESGT's in \cite{GST2,GST5} and certain truncations of the 
$\mathcal{N}=8$
theory \cite{GRW1} make this possibility quite plausible.
A more detailed analysis of the potential and its critical points, 
however, is now 
complicated by the additional scalar potential term $-g^{2}P$ induced by the
tensor field dualization and the additional constraints on the $V_{\ti}$ 
and will therefore be given elsewhere \cite{GZ2}.

\section{ Allowed gauge groups and the
 corresponding representations
of the tensor multiplets}
\setcounter{equation}{0}
In this section we will give a partial classification of the possible gauge
 groups and the representations
under which the tensor fields transform. An attempt at a complete classification
 will be made elsewhere
\cite{GZ2}. 

We will start our discussion of possible gauge groups with the ``Magical''
 supergravity theories defined by simple Jordan algebras of degree 3.

\begin{enumerate}

\item The largest of the magical $\mathcal{N} =2$ supergravity theories is defined by the
exceptional Jordan algebra with the scalar manifold $E_{6(-26)} /F_4$, which we
shall refer to as the exceptional supergravity theory. The exceptional supergravity
theory and its counterparts in four and three dimensions share many of the remarkable
properties of the maximally extended supergravity theories in the respective
 dimensions.  In the exceptional theory one can gauge
the $SO^*(6)=SU(3,1)$ subgroup of the isometry group $E_{6(-26)}$
of the scalar
 manifold while dualizing twelve of the vector fields into tensor fields that
 form a symplectic representation ($6+6$) of $SO^*(6)$. The pure maximal
 Yang-Mills Einstein subsector of this theory is the unique unified
 Yang-Mills Einstein supergravity in five dimensions that was studied in \cite{GST4}.
To gauge a $U(1)_R$ subgroup of the $R$-symmetry $SU(2)_R$ one needs to break the
non-Abelian gauge group $SU(3,1)$ down to a subgroup. One possibility is to gauge
the $U(1)_R$ such that the $SU(3)$ subgroup of $SU(3,1)$ is unbroken. In this case
we obtain a gauged Yang-Mills Einstein supergravity theory with the gauge group
$U(1)_R \times SU(3)$ and 18 tensor multiplets in the symplectic representation
($3+\bar{3}+3+\bar{3}+3+\bar{3}$) of $SU(3)$. The subsector of this theory involving
only 6 tensor multiplets corresponds to the $\mathcal{N} =2$ truncation
 of the gauged $\mathcal{N}=8$
 theory with the gauge group $SU(3)\times U(1)_R$ \cite{GRW1} which admits an AdS
 ground state. 
One can also gauge the subgroup $U(1)_R$ such that one has a vanishing potential
$P^{(R)}$.
 In this case the unbroken non-Abelian symmetry is the $SU(2,1)$ 
subgroup of $SU(3,1)$  with 18 tensor multiplets.

\item
The magical $\mathcal{N} =2$ MESGT defined by the Jordan algebra of $3\times 3$ Hermitian 
matrices over the quaternions has the scalar manifold $SU^*(6)/USp(6)$. One can
 gauge the $SO^*(6)$ subgroup of the isometry group resulting in the unique 
unified Yang-Mills Einstein supergravity \cite{GST4} with no tensor multiplets.
To obtain a Yang-Mills Einstein supergravity with tensor multiplets one has to
gauge a subgroup of $SO^*(6)$. One can gauge the maximal compact subgroup 
$SU(3)\times U(1)$ of $SO^*(6)$ by dualizing 6 of the vector fields to tensor fields
transforming in the symplectic representation ($3+\bar{3}$) of $SU(3)\times U(1)$.   
One can similarly gauge the non-compact subgroup $SU(2,1)\times U(1)$ of $SO^*(6)$.
In both cases one can use the gauge field associated with the Abelian factor to gauge
the $U(1)_R$ symmetry thereby obtaining gauged Yang-Mills/Einstein supergravity
theories with the non-Abelian gauge groups $SU(3)$ and $SU(2,1)$  and six tensor
multiplets, respectively. We expect the generic $SU(2,1)\times U(1)_R$
 gauging to lead to a vanishing potential $P^{(R)}$. 
\item
 The magical MESGT defined by the Jordan algebra of $3\times3$ Hermitian matrices
over the complex numbers has the scalar manifold $SL(3,\mathbf{C})/SU(3)$. In this theory
one can gauge the full compact symmetry group to obtain a Yang-Mills/Einstein
 supergravity theory with the gauge group $SU(3)$ \cite{GST2}. The remaining vector
 field( graviphoton) can be used to gauge the $U(1)_R$ symmetry with a non-vanishing
potential and an AdS ground state. To obtain a Yang-Mills Einstein supergravity
with tensor multiplets one needs to gauge a subgroup of $SU(3)$. One can, for
 example, gauge the $SU(2)\times U(1)$ subgroup while dualizing four of the vector
fields to tensor fields in the symplectic representation ($2+\bar{2}$). One can then
use the graviphoton to obtain a $U(1)_R$ gauged version of this theory. Noncompact
analogs of these theories also exist with $SU(3)$ and $SU(2)$ replaced by $SL(3,\mathbf{R})$
and $SL(2,\mathbf{R})$, respectively.

\item
The smallest of the magical MESGT's has the scalar manifold $SL(3,\mathbf{R})/SO(3)$.
In this case one can gauge the $SL(2,\mathbf{R})$ subgroup of the isometry group while
dualizing two of the vector fields into tensor fields. The remaining vector
field can be used to gauge the $U(1)_R$ symmetry.

\item
For the generic Jordan family  the scalar manifold of the $\mathcal{N}=2$ MESGT is
\begin{equation}
SO(\tn-1,1)\times SO(1,1) / SO(\tn-1)    
\end{equation}

On the other hand the scalar manifold of the generic symmetric non-Jordan family
is of the form
\begin{equation}
SO(\tn,1)/SO(\tn)
\end{equation}
For the latter family, not all the isometries of the scalar manifold can be extended
to symmetries of the Lagrangian \cite{dWvP2}. Only the subgroup $[SO(\tn-1)
\times SO(1,1)] \odot T_{(\tn-1)}$ ( i.e the Euclidean group in $(\tn-1)$ dimensions 
times dilatations ) of $SO(\tn,1)$ extends to a full symmetry of the action.
This can simply be understood by the fact that there is no irreducible symmetric
invariant tensor of rank three of $SO(\tn,1)$.

 One can treat the generic Jordan and 
non-Jordan families in a unified manner as was shown in \cite{GST3}. Consider
a vector ${\bf m}$ in an $\tn$ dimensional Euclidean space with components $m_{i}$.
Then the non-vanishing components of the tensor $C_{\ti\tj\tk}$ can be written
in the form (cf. eq. (\ref{canbasis}))

\begin{eqnarray}
C_{000}=  1               \\ \nonumber
C_{0ij}= -\frac{1}{2} \delta_{ij}                 \\ \nonumber
C_{ijk}=\frac{3}{2} m_{(i} \delta_{jk)}
 -m_{i} m_{j} m_{k}
\end{eqnarray}
where $i, j,..=1,2,..,\tn$.                   
For the generic Jordan family the length squared of the vector ${\bf m}$ is 
two 
\begin{equation}
{\bf m}\cdot {\bf m} =2
\end{equation}
while for the non-Jordan family one has
\begin{equation}
{\bf m}\cdot {\bf m} =\frac{1}{2}.
\end{equation}
It is easy to verify that the above $C_{ijk}$ 
can provide a symplectic representation of {\it only} an Abelian
 subgroup of the compact
symmetry group of $\mathcal{N}=2$ MESGT.
 Therefore, if we are to have tensor fields transforming
nontrivially under the compact gauge group then only products of $U(1)$'s are allowed
in the Yang-Mills Einstein supergravity with tensor multiplets.
 Of course one still has the
option to gauge a non-Abelian subgroup of the compact symmetry group so long
 as the tensor
fields are inert under it. 
\item
The scalar manifolds listed above exhaust the list of $\mathcal{N} =2$ MESGT's 
whose scalar manifolds are symmetric spaces. In addition there is a large
set of other theories whose scalar manifolds admit isometries that extend to
 symmetries of the full action. These include theories whose scalar manifolds
are homogeneous spaces as well as those that are not homogeneous. A complete
list of possible homogeneous spaces  was given in
\cite{dWvP1}. This classification was achieved by showing that the requirement of
 a transitive isometry group allows one to bring  the most general
solution for the symmetric tensor given in the ``canonical basis'' above
 to the form:
\begin{eqnarray}
C_{011}=1 \\ \nonumber
C_{0\bar{\mu}\bar{\nu}}= - \delta_{\bar{\mu}\bar{\nu}} \\ \nonumber
C_{1\bar{i}\bar{j}}= - \delta_{\bar{i}\bar{j}} \\ \nonumber
C_{\bar{\mu}\bar{i}\bar{j}}=\gamma_{\bar{\mu}\bar{i}\bar{j}} 
\end{eqnarray}
where the indices $\ti$ are now split such that $\ti=0,1,\bar{\mu},\bar{i}$ with
$\bar{\mu}=1,2,..,q+1 $ and $\bar{i}=1,2,..,r$. The coefficients
 $\gamma_{\bar{\mu}\bar{i}\bar{j}}$ are ($q+1$) real $r\times r$ matrices
that generate a real Clifford algebra of positive signature
 ${\cal C}(q+1,0)$. The allowed homogeneous (but not symmetric) spaces are, in general, quotients of  ``parabolic groups'' $G$ modded out by their maximal compact
subgroups $H$. The Lie algebra $g$ of the group $G$ is a semi-direct sum:
\begin{equation}
g =   g_0\oplus g_{+1} \end{equation}
\begin{eqnarray}
 g_0 &=& so(1,1)  \oplus so(q+1,1) \oplus {\cal S}_q(P,Q) \nonumber\\
g_{+1} &= &(spinor,vector)\ ,
\end{eqnarray}
where $spinor$ denotes a spinor representation of
$so(q+1,1)$
(of dimension ${\cal D}_{q+1}$) and $vector$ denotes the  vector
representation of ${\cal S}_q(P,Q)$ which is of dimension $(P+Q)$.\footnote{We
 should note that in case the scalar manifold  is a symmetric
space the above Lie algebra gets extended by additional symmetry generators 
belonging to grade $-1$ space transforming in the conjugate representation of
 $ g_{+1}$ with respect to $g_{0}$.} The isotropy group $H$ is 
\begin{equation}
H=SO(q+1)\otimes {\cal S}_q(P,Q)
\end{equation}
The  possible groups ${\cal S}_q(P,Q)$ and the associated real Clifford algebras
were given in \cite{dWvP1} which we list in Table 1.

\begin{table}[htb]
\begin{center}
\begin{tabular}{||c|c|c|l||}\hline
$q$  &${\cal C}(q+1,0)$& ${\cal D}_{q+1}$&${\cal S}_q(P,Q)$
\\ \hline
$-1$ &$\Rbar$    &1         &$SO(P)$     \\
0    &$\Rbar\oplus \Rbar $&1&$SO(P)\otimes SO(Q)$ \\
1    &$\Rbar(2)$ &2         &$SO(P)$     \\
2    &$\Cbar(2)$ &4         &$U(P)$     \\
3    &$\Hbar(2)$ &8         &$USp(2P)$\\
4    &$\Hbar (2)\oplus \Hbar (2)$&8&$USp(2P)\otimes USp(2Q)$\\
5    &$\Hbar(4)$ &16&$USp(2P)$   \\
6    &$\Cbar(8)$ &16&$U(P)$     \\
7    &$\Rbar(16)$&16&$SO(P)$     \\
$n+8$ &  $\Rbar(16)\otimes{\cal C}(n+1,0)$&16 ${\cal D}_n$ &
as for $q=n$\\
\hline
\end{tabular}
\end{center}
\caption{Real Clifford algebras
 ${\cal C}(q+1,0)$.
 $\Rbar$ , 
${\bf C}$
 and $\Hbar$  are the division algebras of real, complex numbers and quaternions,
 respectively,  while
 ${\cal D}_{q+1}$ denotes the
real dimension of an irreducible representation of
the Clifford algebra. The ${\cal S}_q (P,Q)$ is the metric
preserving group in the centralizer of the Clifford algebra
 in the $(P+Q){\cal D}_{q+1}$  dimensional representation.}
\end{table}

Now the gamma matrices $\gamma_{\bar{\mu}\bar{i}\bar{j}}$ provide a symplectic
representation of a  group only for $q=1$ or $q=2$ i.e for $U(1)$ or $SU(2)$.
 Hence one can gauge $SU(2)$
symmetry of the $\mathcal{N}=2$ MESGT for $q=2$ while dualizing the $2P$ vector fields
to tensor fields. One can then use the remaining two $SU(2)$ singlet vector
fields  to gauge the $U(1)_R$ symmetry and/or the Abelian $U(1)$ factor in
$U(P)=U(1)\times SU(P)$. 
For $q=1$ one can gauge the $SO(2,1)$ symmetry  while dualizing
the $2P$ vector fields into tensor fields. The remaining $SO(2,1)$ singlet vector
field can then be used to gauge the $U(1)_R$ symmetry of these theories.
\item
As is clear from  above, the coupling to the tensor fields restricts
the possible non-Abelian symmetry groups greatly for those theories whose
scalar manifolds are symmetric spaces or homogeneous spaces. We would like
to point out that there does exist
 a novel class of (gauged) $\mathcal{N} =2$ Yang-Mills
 Einstein supergravity theories coupled  to tensor multiplets with a rich
set of possible non-Abelian groups that admit symplectic representations.  
 To construct these theories one simply  chooses the {\it arbitrary} tensor
$C_{ijk}$ in the canonical basis (\ref{canbasis}) as follows.
Split the indices $i,j,k..$ as $i=(\bi,M), j=(\bj,N),..$ where
 $\bi,\bj =1,2,..,n-1$ and $M,N,..=1,..,2m$ and identify  
\begin{equation}
C_{\bi\bj\bk}=d_{\bi\bj\bk}
\end{equation}
\begin{equation}
C_{\bi MN}=C_{M\bi N}=C_{MN\bi}=\sqrt{\frac{3}{2}} 
(\Lambda_{\bi})_{MN}=\sqrt{\frac{3}{2}} (\Lambda_{\bi})_{NM}
\end{equation} 
where $d_{\bi\bj\bk}$ are the completely symmetric 
 Gell-Mann $d$-symbols of a Lie group $K$ and the
$(\Lambda_{\bi})$ are the matrices of $2m$ dimensional symplectic
 representation of $K$. Now the $d$-symbols vanish for all simple groups
except for the groups $SU(N), N>2 $ and $Spin(6)$ which is isomorphic to $SU(4)$.
For vanishing $d$-symbols and $m=0$ the cubic form reduces to that of the generic
Jordan family. Thus the theories defined by non-vanishing
$C_{ijk}=d_{ijk}$ can be considered as the non-trivial generalizations of the
generic Jordan family. The first non-trivial example i.e the case of $K=SU(3)$
$d$-symbols ( with $m=0$) 
lead to the magical $\mathcal{N}=2$ MESGT with the scalar manifold
$SL(3,\mathbf{C})/SU(3)$.  The scalar manifold obtained by
taking $C_{ijk}$ to be the $d$-symbols of $SU(N)$ for $N>3$ cannot be a
symmetric or homogeneous space.
 This follows from the fact that for such theories $SU(N)$ act as
 isometries of the scalar manifold that extend to symmetries
of the full Lagrangian. However,it is clear from the list of possible
 homogeneous spaces \cite{dWvP1} that it does not include manifolds with 
such properties. Hence the isometries of the scalar manifolds corresponding
to $C_{ijk}=d_{ijk}$ for $N>3$ in the canonical basis cannot act transitively.
This is perhaps expected from the fact that the number of independent invariants of
a group in its adjoint representation is equal to its rank i.e. the number
of Casimir operators. The term involving $\delta_{ij}$ in the cubic form
corresponds to the quadratic invariant and the term involving $d_{ijk}$ 
corresponds to the third order Casimir. Only for $SU(3)$ do they form
 a complete set of invariants and the resulting scalar manifold is a symmetric
space.  For higher $SU(N)$ ($N>3$) one has invariants of order  up to $N$.

As for the symplectic representations $\Lambda_{\bi}$  of  $SU(N)$, one can ,
 for example, 
choose the reducible
$(N+\bar{N})$ representations corresponding to the standard embedding
 of $U(N)$ in $USp(2N)$ by taking $m=N$ and $\tn = N^2$. 

\end{enumerate}    
\section{Conclusions}
\setcounter{equation}{0}
Our results imply several interesting conclusions.

Whereas the R-symmetry group and the isometry group G of the scalar
manifold $\mathcal{M}$ are entangled with each other for $\mathcal{N}>2$ MESGT's,
and for simple supergravities for  $\mathcal{N}>4$
the case $\mathcal{N}=2$ allows a separate discussion of the gaugings 
of subgroups 
of these two groups. In particular, the issues of the tensor field dualization 
and the gravitino coupling to gauge fields can be completely separated.
It turns out that both mechanisms require their own scalar potential.
The potential due to the introduction of the tensor fields is manifestly 
non-negative and does therefore not admit  an anti-de Sitter solution.
This is in contrast to the pure $U(1)_{R}$-gauging, which involves 
minimal coupling to the gravitini and leads to an indefinite potential
which can sustain  anti-de Sitter vacua. 

Combining both types of gauging, one observes surprisingly little     
interference. 
In particular, the scalar potential is just a sum of the two 
potentials of the individual gaugings. Nevertheless, the analysis of the 
critical points seems   to be more complicated, but is, in general, expected
 to allow anti-de Sitter solutions \cite{GZ2}. A particular example of such
a theory obtained by a truncation of the gauged $N=8$ supergravity does admit an 
AdS vacuum \cite{GRW1}.

The introduction of the tensor fields leads to strong constraints on
the possible gauge groups $K\subset G$ and the representations under which the
tensor fields transform. 
The simultaneous $U(1)_{R}$-gauging further restricts these gauge    
groups $K$,
which is one of the few places where these two types of gaugings interfere 
with each other.
 We gave a list of possible gauge groups and the
corresponding representations of the tensor fields using the known
classification of $\mathcal{N} =2$ MESGT's  whose scalar manifolds are 
 symmetric or homogeneous spaces. We also pointed out the existence of a novel
family of $\mathcal{N} =2$ MESGT's whose scalar manifolds are, in general, not homogenous,
but admit $SU(N)$ isometries. The latter class of theories lead to a richer class
of gaugings with some of the vector fields dualized to tensor fields.

{\bf Acknowledgements:} We would like to thank Sergio Ferrara, Renata Kallosh, 
Raymond Stora and Antoine van Proeyen for useful discussions.

\end{document}